\documentclass[sigconf]{acmart}

\copyrightyear{2025}
\acmYear{2025}
\setcopyright{acmlicensed}
\acmConference[MM '25] {Proceedings of the 33rd ACM International Conference on Multimedia}{October 27--31, 2025}{Dublin, Ireland.}
\acmBooktitle{Proceedings of the 33rd ACM International Conference on Multimedia (MM '25), October 27--31, 2025, Dublin, Ireland}
\acmISBN{979-8-4007-2035-2/2025/10}
\acmDOI{10.1145/3746027.3755035}

\usepackage{booktabs}
\usepackage{multirow}
\usepackage{graphicx}
\usepackage[misc]{ifsym}
\usepackage{color}
\usepackage{hyperref}
\hypersetup{
    colorlinks=true,
    urlcolor=blue, 
    citecolor=green, 
    linkcolor=red 
}

\definecolor{makeRed}{RGB}{237, 28, 36}
\definecolor{itGreen}{RGB}{141, 198, 63}
\definecolor{realBlue}{RGB}{41, 171, 226}

\begin{document}

\title{\raisebox{0cm}{\includegraphics[height=0.5cm]{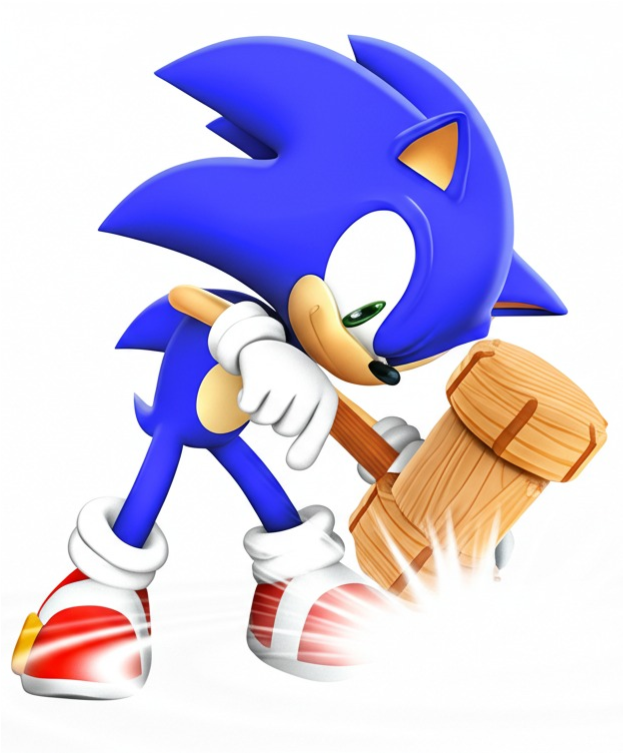}} \textcolor{makeRed}{Sonic}\textcolor{blue}{Gauss}: Position-Aware Physical Sound Synthesis for 3D Gaussian Representations}

\author{Chunshi Wang}
\orcid{0009-0001-5994-2639}
\affiliation{%
  \institution{Zhejiang University}
  \city{Ningbo}
  \state{Zhejiang}
  \country{China}
}
\email{chunshiwang@zju.edu.cn}

\author{Hongxing Li}
\orcid{0009-0001-8310-5814}
\affiliation{%
  \institution{Zhejiang University}
  \city{Ningbo}
  \state{Zhejiang}
  \country{China}
}
\email{hongxing.li@zju.edu.cn} 

\author{Yawei Luo}
\orcid{0000-0002-7037-1806}
\authornote{Corresponding author.}
\affiliation{%
  \institution{Zhejiang University}
  \city{Ningbo}
  \state{Zhejiang}
  \country{China}
}
\email{yaweiluo@zju.edu.cn}

\renewcommand{\shortauthors}{Chunshi Wang, Hongxing Li, \& Yawei Luo}

\begin{abstract}
While 3D Gaussian representations (3DGS) have proven effective for modeling the geometry and appearance of objects, their potential for capturing other physical attributes—such as sound—remains largely unexplored. In this paper, we present a novel framework dubbed \textbf{SonicGauss} for synthesizing impact sounds from 3DGS representations by leveraging their inherent geometric and material properties. Specifically, we integrate a diffusion-based sound synthesis model with a PointTransformer-based feature extractor to infer material characteristics and spatial-acoustic correlations directly from Gaussian ellipsoids. Our approach supports spatially varying sound responses conditioned on impact locations and generalizes across a wide range of object categories. Experiments on the ObjectFolder dataset and real-world recordings demonstrate that our method produces realistic, position-aware auditory feedback. The results highlight the framework's robustness and generalization ability, offering a promising step toward bridging 3D visual representations and interactive sound synthesis. Project page: \textcolor{magenta}{\url{https://chunshi.wang/SonicGauss}}.
\end{abstract}



\begin{teaserfigure}
  \centering
  \includegraphics[width=0.99\linewidth]{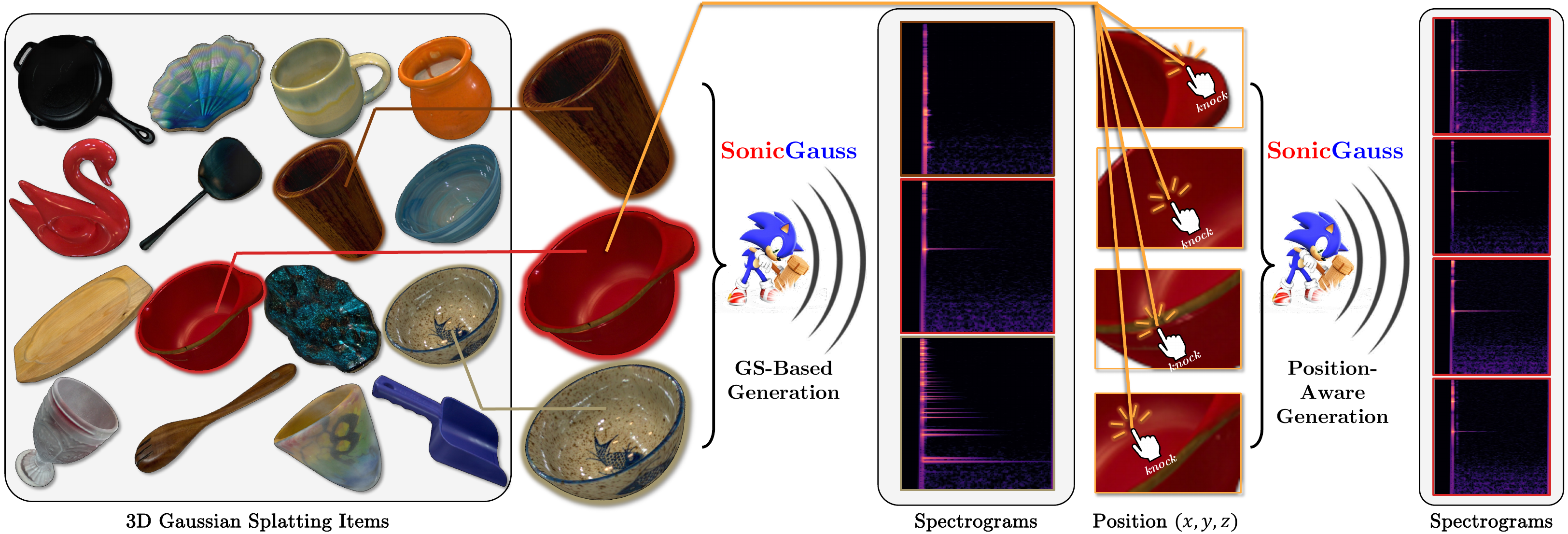}
  \caption{SonicGauss: A novel framework for interactive impact sound synthesis from 3D Gaussian Splatting (3DGS). Our approach bridges the gap between visual and acoustic modalities by directly mapping 3D Gaussian representations to realistic, position-aware sounds. Left: 3D Gaussian Splatting representations of various household objects. Center: Our dual-pathway architecture processes both material properties (via GS-Based Generation) and impact location information (via Position-Aware Generation). Right: The resulting spectrograms demonstrate how our model captures the distinctive acoustic signatures of different materials and positions, enhancing the immersive quality of interactive 3D experiences. }
  \Description{TeaserFigure}
  \label{fig:teaser}
\end{teaserfigure}

\maketitle

\section{Introduction}

Interactive 3D experiences are increasingly integrated into daily life~\cite{ma2025mags,meng2025grounding}, spanning applications such as virtual reality, digital twins, and video games~\cite{quan2025particlegs,luo2024large,miao2025advances}. As these technologies advance, the demand for more immersive and realistic user experiences continues to grow. While significant progress has been made in visual rendering, particularly with methods such as 3D Gaussian Splatting (3DGS)~\cite{3dgs,dy-gs,gs-slam,Wu_2024_CVPR}, the acoustic dimension of these interactions remains comparatively underexplored and underdeveloped. Impact sounds (the sounds produced when objects interact) are particularly crucial for creating a believable sense of presence and material properties within virtual environments. These sounds provide immediate feedback about an object's material properties, structure, and physical response to contact, greatly enhancing user engagement across applications ranging from immersive training simulators to interactive product visualization and gaming.



\begin{figure}
    \centering
    \includegraphics[width=\linewidth]{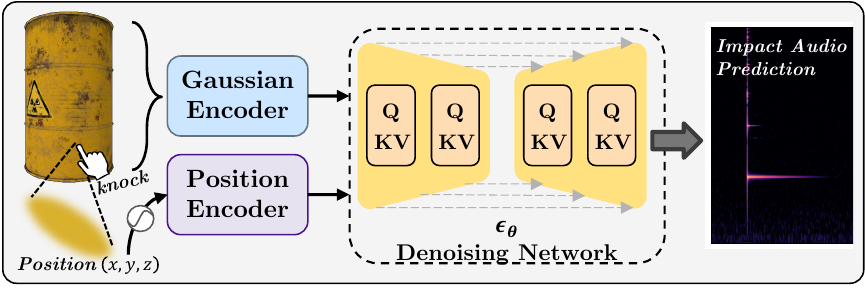}
    \caption{Overview of SonicGauss for interactive impact sound synthesis from 3DGS. Our approach combines a Gaussian Encoder to extract material properties from 3D Gaussian representations with a Position Encoder to capture impact location information. These features are fused and fed into a denoising network to generate position-aware impact audios.}
    \Description{The figure shows a simplified pipeline of SonicGauss with four main components.}
    \label{fig:intro}
\end{figure}

Recent advances in generative models have significantly improved audio synthesis and impact sound generation~\cite{yang2023plugin,gramaccioni2024stable}. However, most existing methods \textbf{rely on intermediate modalities} such as video or explicit material annotations~\cite{zhou2018visual,chen2020generating,liu2025visual}, limiting their applicability in real-time, interactive 3D environments. Furthermore, these approaches often \textbf{assume objects to be acoustically homogeneous}, overlooking the spatial variation in sound responses that arise from different impact locations. Revisiting the physical principles of sound generation reveals that the acoustic behavior of an object is intrinsically linked to its material properties, geometric structure, and the specific site of interaction~\cite{clarke2022diffimpact}. This observation motivates a novel perspective: bypassing intermediate representations and directly mapping visual-geometric information to acoustic attributes—specifically, lifting representations like 3D Gaussian Splatting (3DGS) to the audio domain. While 3DGS provides explicit representations of geometry and appearance and encodes cues potentially indicative of material characteristics, its capacity to inform spatially-aware sound synthesis remains largely unexplored.

In this paper, we present \textbf{SonicGauss}, a novel framework that bridges 3D visual representation and interactive sound synthesis. As illustrated in Fig.\ref{fig:intro}, SonicGauss directly maps 3D Gaussian representations to realistic, position-aware impact sounds. The system comprises two key components: a Gaussian Encoder, which extracts material properties from 3D Gaussian ellipsoids, and a Position Encoder, which encodes the spatial location of the interaction. The outputs of these encoders are fused and input into a denoising diffusion network that transforms noise into plausible acoustic signals. This architecture enables SonicGauss to generate audio that reflects both the material characteristics and spatial acoustic variations of an object. To train the model, we design a three-stage pipeline. First, we fine-tune a text-to-audio diffusion model for impact sound generation using descriptive captions produced by a vision-language model from the ObjectFolder2.0~\cite{objfolder2} dataset. Second, we train the Gaussian Encoder, based on the Point Transformer V3~\cite{ptv3} architecture, to learn material representations via contrastive alignment with textual descriptions. Finally, we incorporate positional encoding and fine-tune the entire system on real-world impact sound recordings from ObjectFolder-Real~\cite{objfolderreal}, enabling accurate and location-sensitive sound generation.

Our contributions can be summarized as follows:

\begin{enumerate}
    \item We present the first framework for generating impact sounds directly from 3DGS representations, enhancing the immersive quality of real-time rendered scenes without requiring intermediate modalities.
    
    \item We introduce a novel architecture that combines a Point Transformer-based Gaussian Encoder with a diffusion-based sound synthesis model, effectively bridging the gap between 3D visual representations and interactive audio.
    
    \item We develop a position encoding mechanism that enables the generation of position-aware sound responses, capturing the subtle acoustic differences that occur when knocking different parts of an object.
    
    \item Through extensive experiments on both synthetic and real-world datasets, we demonstrate the effectiveness of our approach in producing realistic, position-aware impact sounds that generalize across diverse objects.
\end{enumerate}

\section{Related Work}

\subsection{Audio Generation with Diffusion Models}

Recent advances in generative models, particularly latent diffusion approaches, have profoundly transformed multi-modal content development. Diffusion models have exhibited astonishing abilities in producing high-fidelity media across various forms, including imagery~\cite{rombach2022high,podell2023sdxl,hoe2024interactdiffusion,mou2024t2i}, video~\cite{ho2022video,hu2023lamd,chen2023control,ma2025step}, and sound~\cite{kong2020diffwave,shen2023naturalspeech,lee2024voiceldm,copet2023simple,zhang2024instruct}. Their potential to gradually morph noise into organized patterns makes them particularly suitable for audio synthesis, where maintaining intricate temporal designs and frequency traits is critical.

In the domain of text-to-audio generation, models like TangoFlux~\cite{tangoflux} have achieved impressive results with relatively small model sizes (515M parameters) while maintaining generation efficiency. It employs a rectified flow method and presents CLAP-Ranked Preference Optimization (CRPO)~\cite{xu2021crpo} to improve alignment between textual descriptions and synthesized audio. Similarly, AudioLDM~\cite{liu2023audioldm,liu2024audioldm} benefits from a latent diffusion approach to learn compact audio representations, allowing for high-quality generation while decreasing computational demands. These models have significantly advanced the state of the art in general-purpose audio synthesis from text depictions.

Other notable contributions include AudioGen~\cite{kreuk2022audiogen}, which leverages an autoregressive technique for text-to-audio generation, and MusicLM~\cite{agostinelli2023musiclm}, which centers specifically on music generation. While these models excel at generating ambient audios, music, and general audio effects, they are not intentionally designed to capture the unique features of impact sounds linked to explicit 3D objects and materials.

\subsection{Impact Sound Synthesis}

Impact sound synthesis has been approached from various perspectives, ranging from physics-based simulations to learning-based methods. Traditional physics-based approaches~\cite{o2002synthesizing,o2001synthesizing,traer2019perceptually} simulate the vibration modes of objects using techniques such as modal analysis, where the sound is generated by solving the wave equation for a given object geometry and material properties. While physically accurate, these methods require detailed knowledge of material parameters (Young's modulus, Poisson's ratio, damping coefficients) and are computationally expensive, limiting their applicability in interactive scenarios.

With the advent of deep learning, several approaches have emerged to bridge the gap between visual and auditory modalities. Owens et al.~\cite{Owens_2016_CVPR} pioneered the use of recurrent neural networks to generate impact sounds from silent videos, learning the complex mapping between visual dynamics and corresponding acoustic responses. 
Chen et al.~\cite{chen2018visually,chen2020generating} subsequently enhanced this by incorporating perceptual losses and information bottlenecks, yielding audio better aligned with inputs.

Recent developments have focused on combining physics-based knowledge with learning-based methods. DiffSound~\cite{diffsound}, a differentiable sound rendering framework, enables solving inverse difficulties such as estimating parameters and positions by gradient descent. Similarly, physics-driven diffusion models~\cite{phydriven} integrate physical priors with visual information from videos to generate more realistic impact sounds while maintaining interpretability and editability.

The ObjectFolder datasets~\cite{objfolder2,objfolderreal} have been instrumental in advancing multisensory learning, providing both synthetic (ObjectFolder 2.0) and real-world (ObjectFolder-Real) data of household objects with corresponding visual, acoustic, and tactile information. These datasets enable the training of models that can generalize across different objects and sensory modalities, facilitating research in cross-modal synthesis.

\subsection{Limitations of Existing Approaches}

While past efforts have certainly advanced the field, several limitations remain unresolved in current techniques for synthesizing impact sounds. Primarily, most existing methodologies rely on video data or intermediate modalities like images to generate impact sounds, presenting extra intricacy as well as potential sources of error. Additionally, present approaches regularly lack the means to craft position-specific acoustic responses, dealing with objects as homogeneous entities rather than considering the spatial fluctuations in material properties that sway sound formation. Furthermore, numerous techniques necessitate comprehensive material specifications or physics simulations, restricting their extensiveness and real-time functionality.

Moreover, none of the present methodologies directly capitalize on the rich geometric and visual information encoded in 3D Gaussian Splatting representations. As 3DGS rapidly emerges as the preferred technique for real-time rendering of complex scenes, there is a pressing need for approaches that can directly map from this representation to realistic impact noises, rounding out the audiovisual experience for interactive applications.

\begin{figure*}
    \centering
    \includegraphics[width=0.95\linewidth]{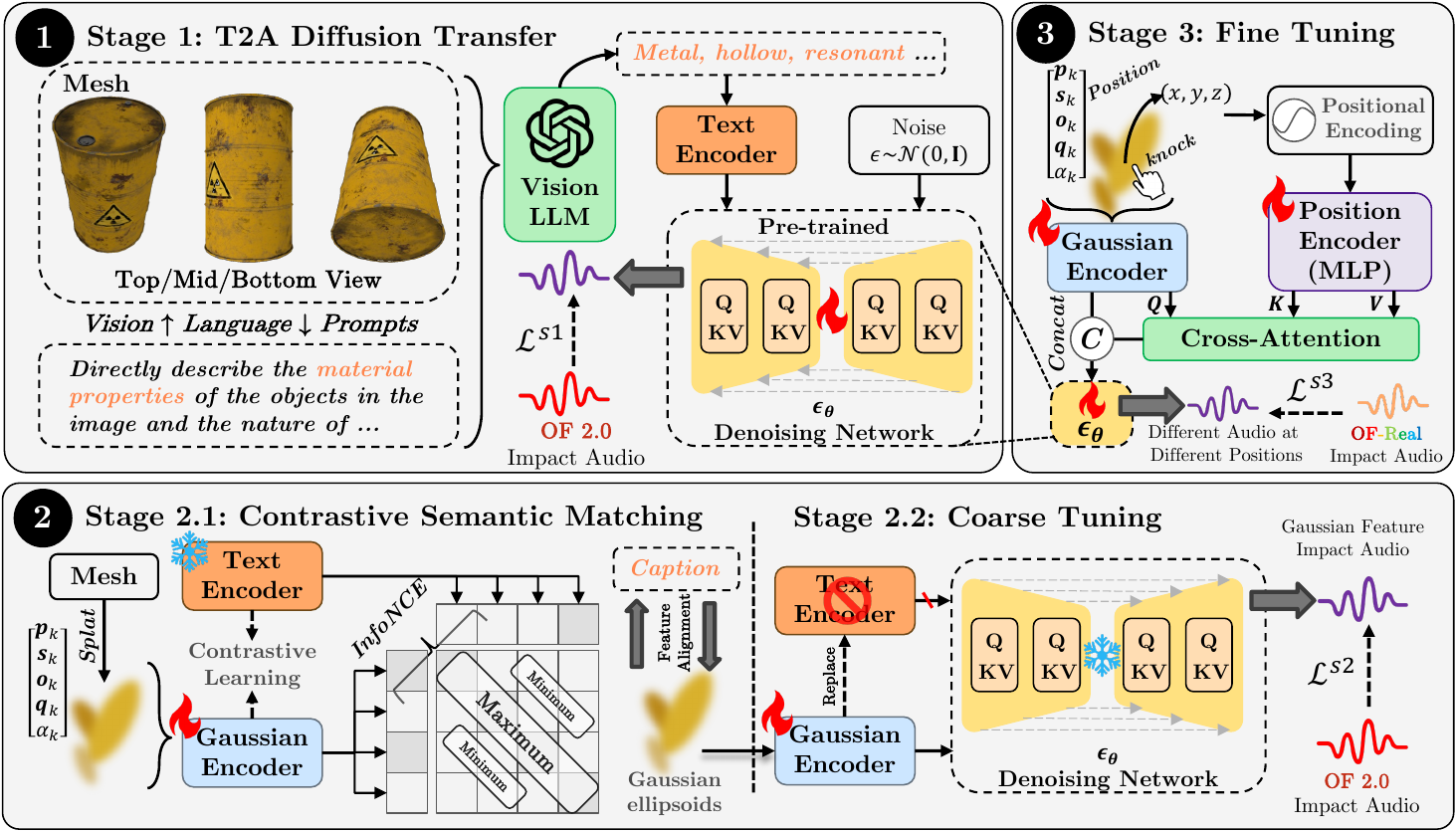}
    \caption{Overview of our SonicGauss framework with its three-stage approach. Stage 1 (T2A Task Transfer): Adapting the TangoFlux diffusion model for impact sound synthesis using Vision-LLM generated material descriptions. Stage 2.1 (Contrastive Semantic Matching): Aligning Gaussian embeddings with text embeddings using InfoNCE loss to transfer material understanding. Stage 2.2 (Coarse Tuning): Replacing the text encoder with the trained Gaussian encoder for direct 3DGS-to-audio generation. Stage 3 (Fine Tuning): Incorporating position-specific responses by introducing a position encoder with frequency embedding and feature fusion with the Gaussian features, trained on real-world impact sounds at different positions.}
    \Description{A diagram showing the three-stage pipeline of SonicGauss. The top panel shows Stage 1 (Task Transfer) with mesh inputs, Vision-LLM, and text-conditioned diffusion model. The bottom left shows Stage 2.1 (Contrastive Semantic Matching) with the Gaussian and text encoders trained with contrastive learning. The bottom right shows Stage 2.2 (Coarse Tuning) with the Gaussian encoder replacing the text encoder. The top right shows Stage 3 (Fine Tuning) with position encoding integration.}
    \label{fig:pipeline}
\end{figure*}

\section{Methodology}

\subsection{Preliminaries}

\textbf{3D Gaussian Splatting (3DGS) }is a popular technique for representing and rendering 3D scenes, offering both high visual quality and real-time performance. This technique directly optimizes a diverse set of parameterized 3D Gaussians to model a scene through a non-neural approach. In contrast to traditional neural radiance fields that leverage neural networks for volumetric modeling, 3DGS represents each point independently as a Gaussian blob. Each Gaussian is parameterized by its position $\boldsymbol{p}_{k} \in \mathbb{R}^3$, scale $\boldsymbol{s}_{k} \in \mathbb{R}^3$, rotation $\boldsymbol{q}_{k}$ (quaternion), opacity $\boldsymbol{o}_{k} \in [0,1)$, and color coefficients $\boldsymbol{\alpha}_{k}$ represented as spherical harmonics.



This representation offers several advantages for our task. First, 3D Gaussians effectively encode both geometric and material attributes in a unified framework. Second, the anisotropic Gaussians' explicit depiction of form supplies natural hints about substance, which are crucial for synthesizing sound. Finally, 3DGS models realize high visual quality while maintaining compact representation, facilitating extraction of meaningful characteristics for cross-modal tasks like sound genesis.

\subsection{Overview}

SonicGauss introduces a novel framework for synthesizing impact sounds from 3D Gaussian Splatting representations. Our method comprises three main stages (Figure~\ref{fig:pipeline}): Text-to-Audio (T2A) Diffusion Transfer, Contrastive Semantic Matching with Coarse Tuning, and Fine Tuning. Each stage addresses specific challenges in generating realistic impact sounds from 3D Gaussian representations, gradually refining the model's ability to produce material-informed and position-reliant audio responses.

Our framework exploits two complementary datasets: ObjectFolder2.0, containing 1,000 objects with impact sounds generated through Finite Element Method (FEM), and ObjectFolder-Real, comprising 100 real-world objects with approximately 35 professionally recorded impact sounds at different places per object. The former is used for initial training and material-aware sound generation, while the latter allows fine-grained position-reliant sound synthesis.

\subsection{Stage 1: Text-to-Audio (T2A) Diffusion Transfer for Impact Sound Synthesis}

The first stage of our pipeline focuses on adapting a pre-trained diffusion model for the specific task of impact sound generation. We build upon TangoFlux~\cite{tangoflux}, a T2A diffusion model originally trained on environmental sounds and general audio descriptions. While TangoFlux has abundant prior knowledge about sound genesis, it requires adjustment to properly synthesize isolated impact sounds rather than continuous audio sequences.

Within the TangoFlux architecture, the denoising network is implemented as a rectified flow model that learns to map noise to realistic audio conditioned on text. Formally, given a noise input $\epsilon \sim \mathcal{N}(0, I)$ and a text condition $c_{text}$, the model estimates a velocity field $v_t$ through:
\begin{equation}
    v_t = u_\theta(x_t, t, c_{text}),
\end{equation}
where $u_\theta$ denotes the velocity network with parameters $\theta$, $x_t$ denotes the state at time $t$, and $t$ is the diffusion time. The output audio $X$ is generated by following the flow from noise to clean audio.

During this task transfer stage, we fine-tune TangoFlux using descriptions of material properties generated by a Vision-LLM(GPT-4o)~\cite{gpt4o}. As shown in Figure~\ref{fig:pipeline}, we initially feed 3D mesh representations from ObjectFolder2.0 into the Vision-LLM, which creates detailed textual descriptions focusing on material properties, such as \textit{``Metallic, hollow, resonant.''} These descriptions, along with the corresponding impact sounds from ObjectFolder2.0, form the training pairs for fine-tuning.

The optimization goal is a flow-matching loss function:
\begin{equation}
    \mathcal{L}^{S1} = \mathbb{E}_{t, x_0, \epsilon} \left[ \| v_t - u_\theta(x_t, t, c_{text}) \|_2^2 \right],
\end{equation}
where $x_0$ denotes the ground truth audio and $x_t = tx_0 + (1-t)\epsilon$ denotes the interpolated state at time $t$.

This stage serves two critical purposes: (1) it shifts the model's focus from general audio generation to specific impact sound synthesis, and (2) it establishes a foundation for mapping textual descriptions of material properties to their corresponding acoustic signatures, which will be leveraged in the subsequent stages of our pipeline.

\subsection{Stage 2: Contrastive Semantic Matching and Coarse Tuning}

The second stage of our approach moves beyond solely textual prompts towards directly processing 3D Gaussian information. This stage is divided into two sub-stages: (1) Contrastive Semantic Matching and (2) Coarse Tuning, as illustrated in Figure~\ref{fig:pipeline}.

\subsubsection{Stage 2.1: Contrastive Semantic Matching}

The key challenge in this stage is establishing a meaningful mapping between 3D Gaussian features and the semantic space of material properties already learned by the text encoder. To address this, we employ a contrastive learning method inspired by CLIP~\cite{clip} (Contrastive Language-Image Pre-training).

We first design a Gaussian Encoder based on Point Transformer V3, which treats 3D Gaussian ellipsoids as multi-channel point clouds. Each Gaussian is represented by its position $\boldsymbol{p}_{k}$, scale $\boldsymbol{s}_{k}$, rotation $\boldsymbol{q}_{k}$, opacity $\boldsymbol{o}_{k}$, and spherical harmonic coefficients $\boldsymbol{\alpha}_{k}$ for appearance. Unlike standard point clouds, 3DGS preserves the rich geometric and appearance information encoded in the gaussian ellipsoids.
Formally, given a collection of Gaussian ellipsoids $G = \{g_1, g_2, ..., g_n\}$ and their corresponding textual descriptions $T = \{t_1, t_2, ..., t_n\}$, we define the contrastive loss as:
\begin{equation}
    \mathcal{L}_{nce} = -\frac{1}{N}\sum_{i=1}^{N} \log \frac{\exp(\text{sim}(f_{\theta_G}(G_i), f_{\theta_T}(T_i))/\tau)}{\sum_{j=1}^{N} \exp(\text{sim}(f_{\theta_G}(G_i), f_{\theta_T}(T_j))/\tau)},
\end{equation}
where $f_{\theta_G}$ is the Gaussian Encoder with parameters $\theta_G$, $f_{\theta_T}$ is the Text Encoder with parameters $\theta_T$, $\text{sim}(\cdot,\cdot)$ denotes cosine similarity, and $\tau$ is a learnable temperature parameter, initialized to 0.07. This loss is designed to drive the Gaussian Encoder so that its outputs are spatially aligned with those of the Text Encoder for the same entity, which will enable semantically meaningful information about materials to be transferred between the two distributions.

\subsubsection{Stage 2.2: Coarse Tuning}

After the contrastive learning phase, we replace the Text Encoder with our trained Gaussian Encoder. The model will now directly process 3D Gaussian inputs, rather than textual descriptions. During this coarse tuning phase, we fine-tune the entire pipeline end-to-end on the ObjectFolder2.0 dataset.

The training loss within this sub-phase is similar to Stage 1, except now it is the Gaussian Encoder that is used for conditioning:
\begin{equation}
    \mathcal{L}^{S2} = \mathbb{E}_{t, x_0, \epsilon} \left[ \| v_t - u_\theta(x_t, t, f_{\theta_G}(G)) \|_2^2 \right],
\end{equation}
where $f_{\theta_G}(G)$ replaces the text condition $c_{text}$ from Stage 1.

This coarse tuning phase thus sets up a one-to-one mapping from 3D Gaussian representations of the world to impact sound synthesis; specifically, it allows our model to extract material properties directly from geometric features. The result is a model capable of generating plausible impact sounds for objects represented by 3DGS, though still without position-specific acoustic responses.

\subsection{Stage 3: Fine Tuning with Position Awareness}

In the third and final stage of our proposed framework, we extend the model's capabilities to generate position-dependent impact sounds. This critical enhancement addresses a key limitation in existing impact sound synthesis approaches—namely, the inability to account for the variation in acoustic responses based on where an object is struck.

\subsubsection{Position Encoder Design}

Inspired by Neural Radiance Fields (NeRF), we incorporate frequency encoding to better represent the position information in our framework. As shown in Figure~\ref{fig:pipeline}, the Position Encoder takes the position of impact $(x,y,z)$ at the object's coordinatesystem and applies the same frequency encoding technique from NeRF to capture high-frequency variation in all parts of space:
\begin{equation}
    \gamma(p) = (\sin(2^0 \pi p), \cos(2^0 \pi p), \cdots, \sin(2^{L-1} \pi p), \cos(2^{L-1} \pi p)),
\end{equation}
where $p = (x, y, z)$ is the impact position and $L=10$ is the number of frequency bands, resulting in a 60-dimensional embedding ($3 \times 2 \times 10$ from the sinusoidal embeddings). The position information then passes through a multi-layer perceptron (MLP) that maps it to a high-dimensional embedding space:
\begin{equation}
    f_{\theta_P}(p) = \text{MLP}([\gamma(p), p]),
\end{equation}
the MLP consists of three fully connected layers with dimensions $[63 \rightarrow 256 \rightarrow 512 \rightarrow d_{joint}]$, where $d_{joint}$ is the joint attention dimension, ReLU activations between layers and $[ \cdot,\cdot ]$ means concatenation.

\subsubsection{Feature Fusion}
The output of the Position Encoder is integrated with the Gaussian Encoder's features through a cross-attention mechanism. This approach allows for dynamic interaction between position information and material properties:
\begin{equation}
    z = \left[f_{\theta_G}(G), \text{CrossAttn}(f_{\theta_G}(G), f_{\theta_P}(p))\right],
\end{equation}
where $\text{CrossAttn}(\cdot,\cdot)$ denotes the cross-attention operation that allows the model to selectively focus on relevant aspects of the position information based on material features. Specifically, we implement this as a standard transformer cross-attention block where Gaussian features serve as queries that attend to position features serving as keys and values, allowing material features to dynamically extract relevant positional context.

\subsubsection{Training on ObjectFolder-Real}

For this stage, we transition from ObjectFolder2.0 to the ObjectFolder-Real dataset, which contains professionally recorded impact sounds from multiple positions on 100 real-world objects. This dataset is crucial for learning the nuanced relationships between impact positions and their resulting sounds.

The training objective for this stage is:
\begin{equation}
    \mathcal{L}^{S3} = \mathbb{E}_{t, x_0, \epsilon} \left[ \| v_t - u_\theta(x_t, t, z) \|_2^2 \right].
\end{equation}

By combining the material awareness from Stage 2 with position-specific responses in Stage 3, our full SonicGauss framework allows interactive synthesis of realistic impact sounds from 3D Gaussian Splatting objects. The result is a framework that can produce varied impact sounds depending on both the type of material in an object and exactly where you hit it.

\section{Experiments}
\subsection{Datasets}

We evaluate SonicGauss on two complementary datasets that span synthetic to real-world impact sounds:

\subsubsection{ObjectFolder 2.0} ObjectFolder 2.0 (OF-2.0)~\cite{objfolder2} has 1,000 common household objects, also as implicit neural representations, collected in a large scale multisensory dataset. The data consists of visual, acoustic, and tactile information about each object. All impact sounds in this dataset are resulting from the Finite Element Method (FEM) simulation, thus achieving physically accurate acoustic responses based on the object material properties. These sounds, however, only reflect one impact per object, and do not capture the acoustic variation that happens with impacts to different locations. This dataset is used in our first two pipeline stages, Task Transfer and Contrastive Semantic Matching with Coarse Tuning, forming a foundation for material-aware sound synthesis.

\subsubsection{ObjectFolder-Real} ObjectFolder-Real (OF-Real)~\cite{objfolderreal} provides the real-world multisensory recordings from 100 household objects that complements the synthetic data. We developed this dataset to meet this demand in a different way than the ObjectFolder 2.0 dataset, as we focused on the important information provided by the set of position-specific acoustic responses, meaning that we recorded close to 35 different impact sounds per object, all created at different knocking positions on the surface of the object. Each impact recording is appended with accurate continuous coordinate information specifying the impacting location on the object's 3D mesh, as well as semi-analytical ground-truth contact force profiles. We use this dataset in the third step of our pipeline Fine Tuning with Position Awareness, which produces spatially varying impact sounds produced by our model transpiring similar to their real world counterparts.

Together, both dataset types enable SonicGauss to measure its performance in material property extraction (from OF-2.0) and position-aware sound generation (from OF-Real), providing a well-rounded evaluation framework for our method of interactive sound synthesis from 3D Gaussian Splatting representations.

\subsection{Data Preprocessing}

To prepare data for SonicGauss, we transformed the original mesh-based object representations into 3D Gaussian Splatting format. Unlike conventional 3DGS workflows that rely on Structure-from-Motion (SfM) for camera parameter estimation and point cloud initialization, we developed a more controlled approach to ensure consistent quality across all objects in our datasets.

We directly used the given mesh geometry and material data for both OF-2.0 and OF-Real. We start the preprocessing pipeline by generating multi-view rendered images for 3DGS optimization. To systematically capture the object, we presented an Orbit Sequence camera trajectory that covers multiple views of the object. Mathematically, for an object whose center of mass is at the origin, the position $\mathbf{c}_i$ of each camera is expressed as:
\begin{equation}
    \mathbf{c}_i = d \cdot 
    \begin{pmatrix} 
        \cos(\phi_e) \sin(\phi_a) \\ \sin(\phi_e) \\ \cos(\phi_e) \cos(\phi_a) 
    \end{pmatrix}, 
\end{equation}
where $d$ is the optimal camera distance determined by the object's bounding box, $\phi_a$ is the azimuth angle sampled at regular intervals $[0^\circ, 15^\circ, 30^\circ, ..., 345^\circ]$, and $\phi_e$ is the elevation angle set at three distinct levels $[-45^\circ, 0^\circ, 45^\circ]$ to capture the object from bottom, horizontal, and top views respectively. This strategic sampling ensures comprehensive coverage of each object's geometry while maintaining consistent viewing conditions across the dataset.
For initialization of 3D Gaussian primitives, we directly sample points from the mesh surface rather than relying on SfM-generated sparse point clouds. This approach preserves the object's detailed geometry and enables more accurate color attribution from the mesh's texture maps. We sample approximately 10k points per object, with density proportional to local surface area, and initialize their colors based on texture sampling at the corresponding UV coordinates. We give more details about data preprocessing in the appendix.


For audio preprocessing, we standardize all impact sound recordings to a uniform sampling rate of 16k and duration of 3 seconds.

\subsection{Implementation Details}

All models implemented in this framework are built with PyTorch. The text-to-audio component of SonicGauss is derived from the original TangoFlux due to its existing capability of generating audio. For the Gaussian Encoder, we reuse the pre-trained Point Transformer V3 from SplatFormer, but not the decoder part (the upsampling part of the U-shaped structure), and only keep the encoding layers. We introduce a max pooling operation at the bottleneck to extract overall Gaussian features that imply material properties. During training, we utilize the same hyperparameters for all three stages of our pipeline. We train each stage for 80 epochs with a batchsize of 2, where AdamW with a learning rate of $1 \times 10^{-4}$ is the optimizer used. For training stability and convergence, we use a cosine learning rate schedule with warmup. To standardize toward established CLIP-style contrastive learning practices, we set the batchsize to 8 for the contrastive semantic matching in Stage 2.1 and employ the InfoNCE loss to supervise the feature alignment of the Gaussian and text embeddings. All experiments were performed in float32 precision on 4 NVIDIA RTX 4090 GPUs. We use the OpenAI GPT-4o model as our caption generation model.

\subsection{Evaluation Details}

In this section, we employ quantifiable metrics with orthogonal evaluation criteria to assess the quality and fidelity of the generated impact sounds. Fréchet Audio Distance (FAD) reflects the statistical closeness between two distributions of audio features based on the matched frame size. Lower FAD scores mean higher quality and more resemblance to a real sound. KL Sigmoid computes the Kullback-Leibler divergence between spectral features of reference and generated audio, normalizing them via a sigmoid function to obtain an interpretable score between 0 and 1, where higher scores imply better performance.
We also used the Inception Score Mean (IS-Mean) and Standard Deviation (IS-Std) adapted for audio evaluation. IS-Mean jointly assesses both quality and diversity of generated samples through a pre-trained audio classification network, with higher scores indicating better overall generation quality. IS-Std complements this by measuring consistency across generated samples, where lower values suggest more stable output quality across different conditions. Together, these metrics provide a comprehensive evaluation framework that aligns well with human perceptual judgment of impact sound quality.


Although several prior works have explored impact sound generation from videos (e.g.,~\cite{Owens_2016_CVPR}), we do not conduct a direct comparison with these methods due to fundamental differences in datasets and task paradigms. Specifically, the video-based dataset used in~\cite{Owens_2016_CVPR} features a markedly different material distribution, including numerous outdoor elements such as clothing, plastic bags, synthetic rocks, and paper—materials absent from the ObjectFolder datasets employed in our study. Furthermore, video-based approaches often regard object-object interactions captured from fixed viewpoints, limiting their comparability to our 3D Gaussian-based method, which enables viewpoint-aware synthesis. These discrepancies introduce significant confounding variables, rendering direct quantitative comparisons potentially misleading and not reflective of true methodological differences.

\subsection{Quantitative Results}

\begin{table}
    \caption{Quantitative evaluation of SonicGauss across different pipeline stages.}
    \label{tab:main_results}\begin{tabular}{@{}c|c|cccc@{}}
    \toprule
    \multicolumn{1}{l|}{Dataset} & Stage & FAD↓            & KL Sig↓         & IS Avg.↑        & IS Std.↓          \\ \midrule
    \multirow{2}{*}{OF-2.0}     & 1     & 1.6848      
        & 0.3442          & 1.0221          & 1.93$ \times 10^{-3}$          \\
                                & 2     & \textbf{1.1050} & \textbf{0.3930} & \textbf{1.0769} & \textbf{1.82$ \times 10^{-3}$} \\ \midrule
    OF-Real                     & 3     & \textbf{0.7298} & \textbf{0.2068} & \textbf{1.0133} & \textbf{2.90$ \times 10^{-4}$} \\ \bottomrule
    \end{tabular}
\end{table}

Table~\ref{tab:main_results} presents a comprehensive quantitative evaluation of SonicGauss across different pipeline stages and datasets. Our results demonstrate consistent improvement as the model progresses through each stage of our proposed pipeline.

For OF-2.0, we observe a significant performance gain when transitioning from Stage 1 (T2A Diffusion Transfer) to Stage 2 (Contrastive Semantic Matching with Coarse Tuning). Specifically, the FAD score improves from 1.6848 to 1.1050, indicating that the generated audio distributions become substantially closer to the reference distribution. Similarly, we see improvements in the KL Sigmoid score (0.3442 to 0.3930), IS Average (1.0221 to 1.0769), and IS std ($1.93 \times 10^{-3}$ to $1.82 \times 10^{-3}$). These consistent improvements across all metrics validate the effectiveness of our contrastive semantic matching approach in better aligning Gaussian and text embeddings, resulting in higher quality impact sound generation.

For OF-Real, our Stage 3 (Fine Tuning with Position Awareness) achieves the best overall performance with an FAD score of 0.7298. This substantial improvement demonstrates that incorporating position-specific acoustic information significantly enhances the model's ability to synthesize realistic impact sounds. The KL Sigmoid score of 0.2068 further confirms the spectral fidelity of our generated samples compared to the ground truth. While the IS Average (1.0133) is slightly lower than Stage 2, this is expected as OF-Real contains more challenging real-world recordings with higher acoustic complexity and variability.
These quantitative results validate our hypothesis that a staged training approach with progressive refinement from material-aware to position-aware sound synthesis leads to superior performance in interactive sound generation for 3D Gaussian representations.
\begin{figure*}[ht]
    \centering
    \includegraphics[width=0.9\linewidth]{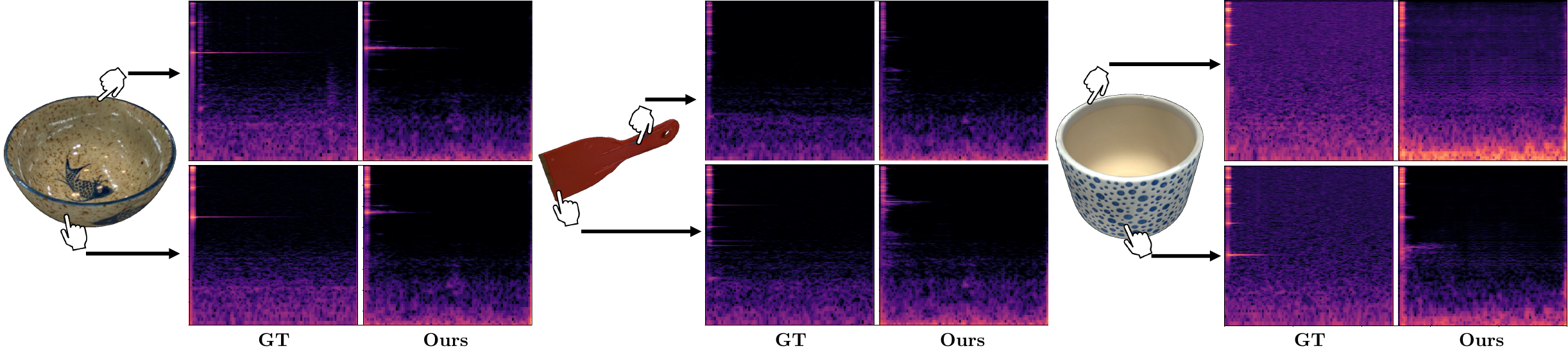}
    \caption{Comparison between ground truth (GT) and SonicGauss-generated spectrograms for impact sounds on various household objects. For each object, we show spectrograms from impacts at different positions.}
    \Description{The figure shows spectrograms comparing ground truth and generated impact sounds for various household objects struck at different positions.}
    \label{fig:results_big}
\end{figure*}

\subsection{Human Perceptual Evaluation}

To complement our objective metrics, we conducted a human perceptual evaluation to assess the perceived quality and realism of our generated impact sounds. Participants evaluated material matching accuracy, sound quality, and position-specific variation.
\begin{table}[h]
    \caption{Human perceptual evaluation results (score range: 0-1, higher is better). DA: Direct Average, WA: Weighted Average.}
    \label{tab:human_eval}
    \resizebox{\linewidth}{!}{
    \begin{tabular}{@{}l|cc|cc|cc@{}}
    \toprule
    \multirow{2}{*}{Dataset} & \multicolumn{2}{c|}{Material Matching} & \multicolumn{2}{c|}{Sound Quality} & \multicolumn{2}{c}{Position Matching} \\
     & DA & WA & DA & WA & DA & WA \\ \midrule
    OF-2.0 & 0.844 & 0.844 & 0.858 & 0.856 & - & - \\
    OF-Real & 0.906 & 0.908 & 0.895 & 0.895 & 0.837 & 0.842 \\ \midrule
    Overall & \multicolumn{2}{c|}{0.875} & \multicolumn{2}{c|}{0.876} & \multicolumn{2}{c}{0.839} \\ \bottomrule
    \end{tabular}
    }
\end{table}

Table~\ref{tab:human_eval} presents results from our human evaluation study. For this evaluation, we randomly sampled unseen data from the test set and distributed them across three separate questionnaires. The Direct Average (DA) represents the average of each questionnaire's mean scores, while the Weighted Average (WA) weights each questionnaire by its number of participants. The results demonstrate strong perceptual quality across all dimensions. For material matching, our model achieved scores of 0.844 and 0.908 on OF-2.0 and OF-Real datasets respectively, indicating that participants found the generated sounds appropriately reflected the material properties visible in the 3D representations. Sound quality ratings were similarly high (0.856 and 0.895), suggesting the generated audio was perceived as realistic and natural-sounding. The position matching score of 0.842 for OF-Real further validates our model's ability to generate position-specific acoustic responses that align with human expectations. Notably, performance on the OF-Real dataset consistently exceeded that on OF-2.0 across all metrics, aligning with our quantitative findings and suggesting that the fine-tuning stage with real-world recordings significantly enhances perceptual quality.

\subsection{Qualitative Results}
To complement our quantitative evaluation, we present qualitative results through spectrogram comparisons between ground truth and SonicGauss-generated impact sounds. Figure~\ref{fig:results_big} showcases spectrograms from multiple objects in the OF-Real dataset, demonstrating our model's ability to synthesize material-appropriate impact sounds that vary realistically with impact position.
\begin{figure}[ht]
    \centering
    \includegraphics[width=0.95\linewidth]{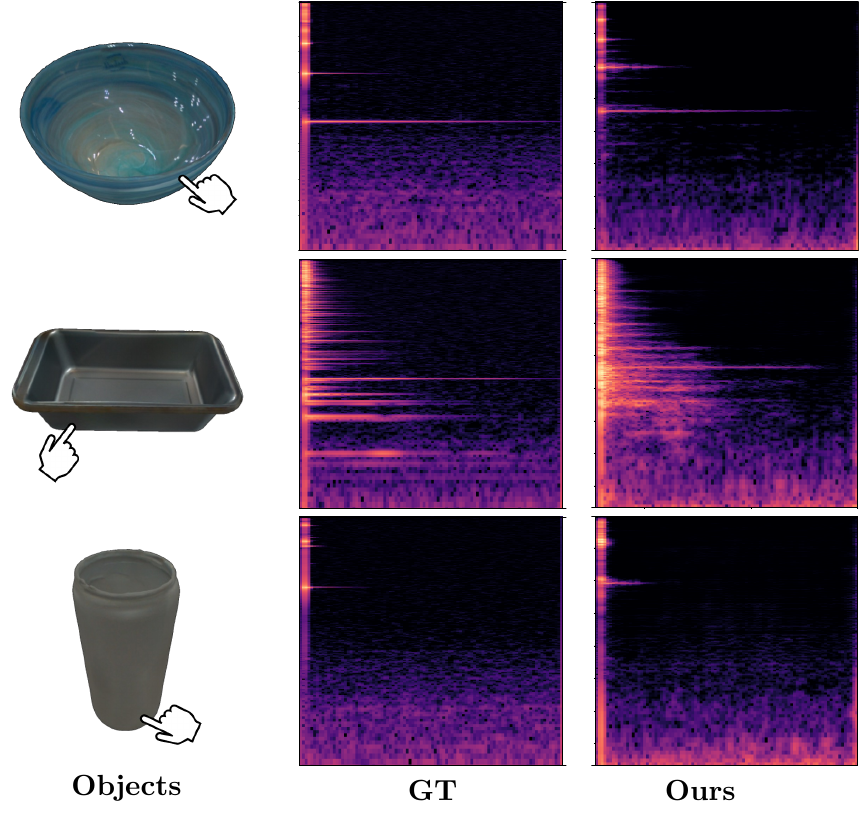}
    \caption{Detailed spectrogram analysis comparing ground truth (GT) and our generated impact sounds for three objects with distinct material properties: a ceramic bowl, a metal baking pan, and a plastic cup.}
    \Description{The figure shows detailed spectrogram comparisons between ground truth and generated impact sounds for three objects made of different materials.}
    \label{fig:results_small}
\end{figure}

Figure~\ref{fig:results_small} presents a more detailed examination of our model's performance across objects with distinctly different material properties. The ceramic bowl (top row) exhibits a characteristic diffuse mid-frequency response with moderate decay times, which our model reproduces with high fidelity. The metal baking pan (middle row) presents a particularly challenging case with its complex resonant structure featuring prominent harmonics and extended decay times. SonicGauss successfully captures these distinctive spectral patterns, though with slightly increased energy in the mid-frequency range. For the plastic cup (bottom row), our model accurately reproduces the dampened response with limited harmonic content typical of plastic materials. \textbf{We encourage readers to explore the interactive sound generation examples provided in our supplementary material to more effectively assess the performance and realism of SonicGauss.}

\subsection{Ablation on Feature Fusion Strategies}
\begin{table}
    \caption{Ablation study on different feature fusion strategies. $\Delta$ shows the absolute improvement of cross-attention over concatenation.}
    \label{tab:ablation_fusion}
    \begin{tabular}{@{}l|cccc@{}}
    \toprule
    Fusion Type & FAD↓   & KL Sig.↓ & IS Avg.↑ & IS Std.↓ \\ \midrule
    Concat      & 0.8666 & 0.3303   & 1.0127   & 3.16$\times 10^{-4}$ \\
    \textbf{$+$CrossAttn} & \textbf{0.7298} & \textbf{0.2068} & \textbf{1.0133} & \textbf{2.90$\times 10^{-4}$} \\ \midrule
    \multicolumn{1}{c|}{\textcolor{gray}{\textbf{$\Delta$ Gain}}} & \textcolor{gray}{\textbf{0.1368}} & \textcolor{gray}{\textbf{0.1235}} & \textcolor{gray}{\textbf{0.0005}} & \textcolor{gray}{\textbf{2.60$\times 10^{-5}$}} \\ \bottomrule
    \end{tabular}%
\end{table}
To investigate the impact of different feature fusion strategies, we conducted an ablation study comparing simple concatenation with our proposed cross-attention mechanism. As shown in Table~\ref{tab:ablation_fusion}, directly concatenating Gaussian and positional features ($z = [f_{\theta_G}(G), f_{\theta_P}(p)]$) provides a reasonable baseline but fails to capture the nuanced interactions between material properties and impact positions. In contrast, our cross-attention approach, where Gaussian features attend to positional information, achieves significantly better performance across all metrics.

\subsection{Ablation on Contrastive Learning Strategy}

\begin{table}
    \caption{Ablation study on the impact of InfoNCE loss in Stage 2.1. Results demonstrate how contrastive learning improves the quality of generated sounds in Stage 2.}
    \label{tab:ablation_infonce}
    \begin{tabular}{@{}l|cccc@{}}
    \toprule
    Method & FAD↓ & KL Sig↓ & IS Avg.↑ & IS Std.↓ \\ \midrule
    w/o InfoNCE & 1.2377 & 0.4076 & 1.0169 & 1.86$\times 10^{-3}$ \\
    \textbf{w/ InfoNCE} & \textbf{1.1050} & \textbf{0.3930} & \textbf{1.0769} & \textbf{1.82$\times 10^{-3}$} \\ \midrule
    \multicolumn{1}{c|}{\textcolor{gray}{\textbf{$\Delta$ Gain}}} & \textcolor{gray}{\textbf{0.1327}} & \textcolor{gray}{\textbf{0.0146}} & \textcolor{gray}{\textbf{0.0600}} & \textcolor{gray}{\textbf{4.00$\times 10^{-5}$}} \\ \bottomrule
    \end{tabular}%
    \vspace{-10pt}
\end{table}

To validate the effectiveness of our contrastive learning approach in Stage 2.1, we conducted an ablation study comparing models trained with and without the InfoNCE loss during the Contrastive Semantic Matching phase. As shown in Table~\ref{tab:ablation_infonce}, incorporating InfoNCE loss significantly improves the model's performance across all metrics. The FAD score decreases from 1.2377 to 1.1050, indicating a closer match to the reference audio distribution. Similarly, the Inception Score mean shows a substantial improvement from 1.0169 to 1.0769, suggesting higher quality and diversity in the generated samples.
These results confirm that the contrastive learning phase effectively serves as a bridge that allows our model to leverage the rich semantic understanding of materials already present in the text encoder, significantly enhancing the quality of generated sounds.

\section{Conclusion}

In this paper, we introduced SonicGauss, a novel framework that directly lifts 3DGS representations to enable the position-aware impact sounds. Our three-stage pipeline (comprising T2A Diffusion Transfer, Contrastive Semantic Matching with Coarse Tuning, and Fine Tuning with Position Awareness) effectively extracts material properties from 3DGS representations and generates spatially varying acoustic responses. Through comprehensive quantitative and perceptual evaluations on both synthetic and real-world datasets, we demonstrated that our approach produces realistic impact sounds that accurately reflect material characteristics and positional acoustic variations.

To the best of our knowledge, this is the first work to explore the potential of 3D Gaussian Splatting representations for acoustic synthesis. We hope this study will also inspire future investigations into other cross-modal physical properties embedded within the 3DGS framework.

\section*{Acknowledgment}
This work was supported by the National Natural Science Foundation of China (62293554, U2336212), National Key R\&D Program of China (SQ2023AAA01005), "Pioneer" and "Leading Goose" R\&D Program of Zhejiang (2024C01073), Natural Science Foundation of Zhejiang Province, China (LZ24F020002), Ningbo Innovation "Yongjiang 2035" Key Research and Development Programme (2024Z292), and Young Elite Scientists Sponsorship Program by CAST (2023QNRC001).

\bibliographystyle{ACM-Reference-Format}
\bibliography{sample-base}

\end{document}